# Quantum anomalous Hall effect in ferromagnetic transition metal halides

Chengxi Huang[1,2] Jian Zhou[2,*] Haiping Wu[1], Kaiming Deng[1], Puru Jena[2,*], Erjun Kan[1,*]

[1] *Department of Applied Physics and Key Laboratory of Soft Chemistry and Functional Materials (Ministry of Education), Nanjing University of Science and Technology, Nanjing, Jiangsu 210094, P. R. China*

[2] *Department of Physics, Virginia Commonwealth University, Richmond, Virginia 23284, United States*

Abstract

The quantum anomalous Hall (QAH) effect is a novel topological spintronic phenomenon arising from inherent magnetization and spin-orbit coupling. Various theoretical and experimental efforts have been devoted in search of intrinsic QAH insulators. However, up to now, it has only been observed in Cr or V doped $(Bi,Sb)_2Te_3$ film in experiments with very low working temperature. Based on the successful synthesis of transition metal halides, we use first-principles calculations to predict that $RuI_3$ monolayer is an intrinsic ferromagnetic QAH insulator with a topologically nontrivial global band gap of 11 meV. This topologically nontrivial band gap at the Fermi level is due to its crystal symmetry, thus the QAH effect is robust. Its Curie temperature, estimated to be ~360 K using Monte-Carlo simulation, is above room temperature and higher than most of two-dimensional ferromagnetic thin films. Inclusion of Hubbard *U* in the Ru-*d* electrons does not affect this result. We also discuss the manipulation of its exchange energy and nontrivial band gap by applying in-plane strain. Our work adds a new experimentally feasible member to the QAH insulator family, which is expected to have broad applications in nanoelectronics and spintronics.

---
* Emails: J.Z.: jzhou2@vcu.edu; P.J.: pjena@vcu.edu; E.K.: ekan@njust.edu.cn

**I. Introduction**

The discovery of topological insulators (TIs) is one of the most important developments in condensed matter physics during the last decade [1-4]. With its bulk being semiconducting, the edge of a two-dimensional (2D) TI is metallic, showing quantum spin Hall conductivity, protected by time reversal symmetry. An interesting alternate topological quantum matter, quantum anomalous Hall (QAH) effect, arises when the time reversal symmetry is broken intrinsically, usually induced by internal magnetism [5-7]. This was first predicted by Haldane [8]. Subsequently, some 2D materials, such as transition metal (TM) doped TIs [9-11], TM decorated graphene [12,13], Rashba spin-orbit coupling and exchange field induced silicene [14,15], TM based organometallic frameworks [16,17], heavy element layers [18], *p*-band optical systems [19], noncollinear antiferromagnetic $K_{0.5}RhO_2$ layer [20], and semi-functionalized stanene or germanene [21], are theoretically predicted to possess QAH effect. In these materials, the spin-orbit coupling (SOC) opens a global band gap at the Fermi level, resulting in topologically nontrivial insulating property. These QAH insulators are also referred to as Chern insulators, as their topological invariant Chern number is nonzero. In spite of being insulating in the bulk, the QAH insulators feature dissipation-less metallic chiral edge states with quantized conductivity, which makes them appealing for high efficiency quantum devices and spintronic applications.

Up to now, the QAH effect has only been experimentally observed in Cr or V doped $(Bi,Sb)_2Te_3$ thin film at very low operation temperature (< 85 mK), and the QAH conductance completely vanishes at 2 K [22-24]. For practical interests, one important challenge in synthesizing QAH insulators is to control the distribution of TM atoms, so that weakening of SOC by charge inhomogeneity can be diminished [25,26]. In addition, the synthesis of such thin film is based on molecular beam epitaxy which is expensive and difficult to manipulate. Therefore, search and design of robust and experimentally feasible QAH insulators is important and still ongoing.

Recently, the experimentally synthesized TM[III] halides [27,28] have received much attention due to their potential applications in spintronics [29]. Due to the weak interlayer van der Waals interactions, these 3D layered crystals can be easily exfoliated down to 2D monolayers [30-32], where the TM atoms are uniformly distributed in a honeycomb structure. While most TM[III] halide monolayers are discovered to be normal metal or semiconductors [31-33], in this study we find that the ferromagnetic (FM) ruthenium halide ($RuX_3$, X=Cl, Br, I) monolayers hold the possibility of being topologically nontrivial. Note that previous experimental and theoretical studies have shown that large halogen ligand or in-plane tensile strain can stabilize their FM coupling against antiferromagnetic (AFM) configuration [31,32]. Besides, the SOC effect increases in heavier elements. Hence, here we use $RuI_3$ monolayer as an exemplary material to study their electronic and magnetic properties by using first-principles calculations. Our results reveal that the ground state of $RuI_3$ monolayer is FM with estimated Curie temperature $T_c$ to be above the room temperature (~360 K). *Ab initio* molecular dynamics (AIMD) simulations confirm its thermal stability at 500 K. A clear Dirac cone in the spin down channel appears at the *K* point in the Brillouin zone near the Fermi level of its band structure. This Dirac cone, due to hybridization of ligand field induced spin down Ru-*e* orbitals, is protected by the real space inversion symmetry of the Ru sublattice. After including SOC interactions, the Dirac cone opens a local band gap of 103 meV, showing a topologically nontrivial feature. The system becomes an insulator with global band gap of 11 meV, in which QAH conductance appears. Thus, we predict that the $RuI_3$ monolayer is an intrinsic QAH insulator. This QAH effect is robust against any perturbation that keeps the crystal symmetry. The FM configurations of $RuCl_3$ and $RuBr_3$ monolayers are also discussed, where we find similar topologically nontrivial characters at *K*.

**II. Computational Details**

Our first-principles calculations are based on spin polarized density functional theory (DFT) with generalized gradient approximation (GGA) for exchange-correlation potential given by Perdew, Burke, and Ernzerhof (PBE) [34] as implemented in the Vienna *Ab initio* Simulation Package (VASP) [35]. A vacuum space of 20 Å along the *z* direction was adopted to model the 2D system. The projector augmented wave (PAW) method [36] was used to treat the core electrons, while the valence electrons were represented using planewave basis set. The planewave cutoff energy was set to be 500 eV, and the first Brillouin zone was sampled using a Γ-centered (12×12×1) Monkhorst-Pack grid [37]. The convergence criteria for energy and force were set to be $10^{-5}$ eV and 0.01 eV/Å, respectively. The SOC was included in the self-consistent calculations. In order to integrate Berry curvature, a much denser *k*-mesh of (120×120×1) was adopted. To verify the GGA results, we also repeated our calculations using the GGA+*U* method [38], with effective Hubbard *U* value of 0.5, 1.0, and 1.5 eV for Ru-*d* electrons. Very similar results have been obtained (Table S2 in Supplemental Material [39]). We fit a tight-binding Hamiltonian by using maximally localized Wannier functions (MLWFs) [40] to the DFT calculated bands, as implemented in the Wannier90 package [41].

### III. Results

#### A. Structure and magnetic property of RuI$_3$ monolayer

Figure 1(a) shows the optimized structure of the RuI$_3$ monolayer which consists of three flat atomic layers: top-I, middle-Ru, and bottom-I layer. The equilibrium lattice constant of hexagonal lattice is 7.10 Å, larger than that of RuCl$_3$ monolayer (5.96 Å [42]). Each Ru atom is coordinated to six I atoms with Ru-I bond length of 2.71 Å. The geometric structure is crystallographically subject to the *P*-31*m* layer group (no. 71). The "thickness" of this monolayer, defined as the distance between the vertical coordinates of the top-I layer and the bottom-I layer, is 3.05 Å. We also calculate its formation energy $E_f = (E_{\text{RuI3}} - 1/4\mu_{\text{Ru}} - 3/4\mu_{\text{I}})$, where $E_{\text{RuI3}}$ is the cohesive energy of RuI$_3$ monolayer. The chemical potential $\mu_{\text{Ru}}$ and $\mu_I$ are taken from the cohesive energy of hcp Ru crystal and I$_2$ molecule, respectively. The calculated formation energy of RuI$_3$ monolayer is –0.23 eV/atom. This negative value is indicative of exothermic reaction. The thermal stability is examined by performing AIMD simulations up to 500 K (Figure S1 in [39]), which implying that the exfoliation reaction to obtain RuI$_3$ monolayers can be carried out at high temperature.

Next, we explore the electronic and magnetic properties of RuI$_3$ monolayer. Since each I atom needs one electron from Ru (with its valence state of $4d^75s^1$), the formal oxidation state of Ru is +3. There leave five $d$ electrons on each Ru atom, and our calculation shows that each Ru atom carries ~1 $\mu_B$ magnetic moment. In order to determine the optimal magnetic coupling, we consider four possible magnetic configurations (one FM and three AFM) as shown in Fig. 1(b). We find that the FM state has the lowest total energy [spin density shown in Figure 1(c)]. The relative energies between the FM and AFM states are listed in Table S1 [39]. During our calculation the Néel-AFM configuration always automatically converged to nonmagnetic state, whose total energy is higher than that of the FM state by 42 meV per formula unit (RuI$_3$, denoted as f.u. thereafter). The zigzag-AFM and stripy-AFM states are energetically higher than that of the FM coupling by 20 and 36 meV/f.u., respectively. The FM coupling remains stable when Hubbard $U$ correction is included on the Ru-$d$ electron, but the exchange energy reduces with $U$ (Table S2 in [39] and Ref. 11). With Hubbard $U$ = 1.5 eV, the zigzag-AFM state lies higher than the FM state by 12 meV. This indicates that the estimated Curie temperature will be reduced to ~60%, which should still be observable experimentally under high enough temperature.

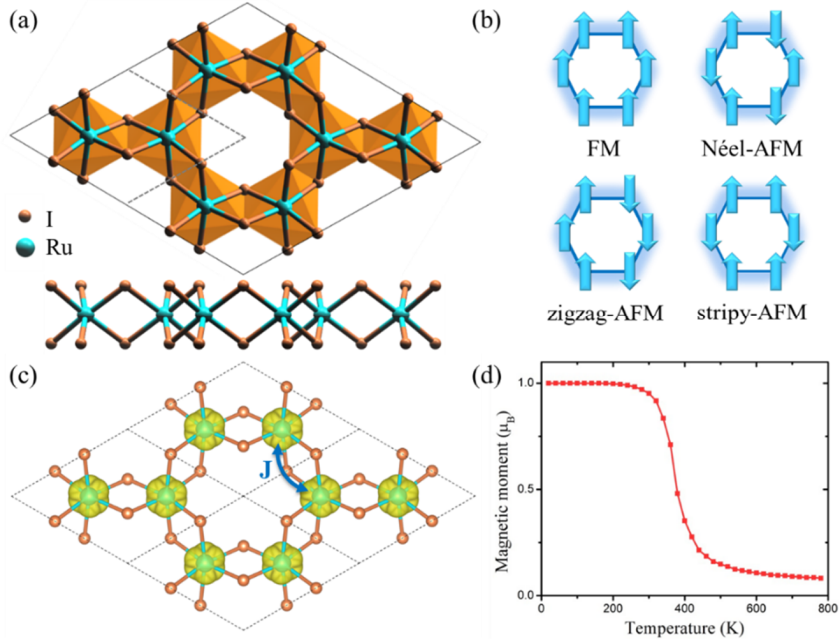

FIG. 1. (a) Top and side view of the optimized 2D RuI$_3$ monolayer. Dashed rhombus refers to the unit cell. (b) Different magnetic configurations. (c) Spin density (iso-value of 0.04 $e$/Å$^3$) and exchange path $J$. (d) Magnetic moment per formula unit as a function of temperature from Monte-Carlo simulation.

To examine the spin dynamical stability against temperature, we use Ising model to describe the spin Hamiltonian, i.e. $H = -\sum_{\langle ij \rangle} J S_i \cdot S_j$, where $J$ refers to the nearest-neighbor exchange parameter [Fig. 1(c)], $S = ½$ according to our calculation, and the summation runs over all nearest-neighbor Ru. The $J$ is calculated to be 82 meV, with positive value indicating FM exchange coupling. We perform a Monte-Carlo simulation to estimate its Curie temperature ($T_c$). A (20×20) supercell is adopted to reduce translational constraint. The magnetic moment per f.u. is taken after the system reaches equilibrium state at a given temperature. In Figure 1d, we see that $T_c$ is ~360 K, which is above room temperature and higher than those of most 2D FM nanomaterials [43-45].

**B. Band structure without including spin-orbit coupling (SOC)**

To gain insight into the electronic properties of FM RuI$_3$ monolayer we calculate the electronic band structure and projected density of states (PDOS). Figure 2a shows the spin-polarized band structure of the FM ground state without including SOC. We find two Dirac cones at the $K$ point in the spin down channel, denoted as $D_K^\downarrow$ and $D_K^{\downarrow\prime}$. The $D_K^\downarrow$ is located slightly above the Fermi level ($E_F$+5 meV with $E_F$ the Fermi energy), and the $D_K^{\downarrow\prime}$ is below the Fermi level ($E_F$–265 meV). From the PDOS (Figure 2b), we see that both these Dirac cones are mainly contributed by Ru-$d$ orbitals.

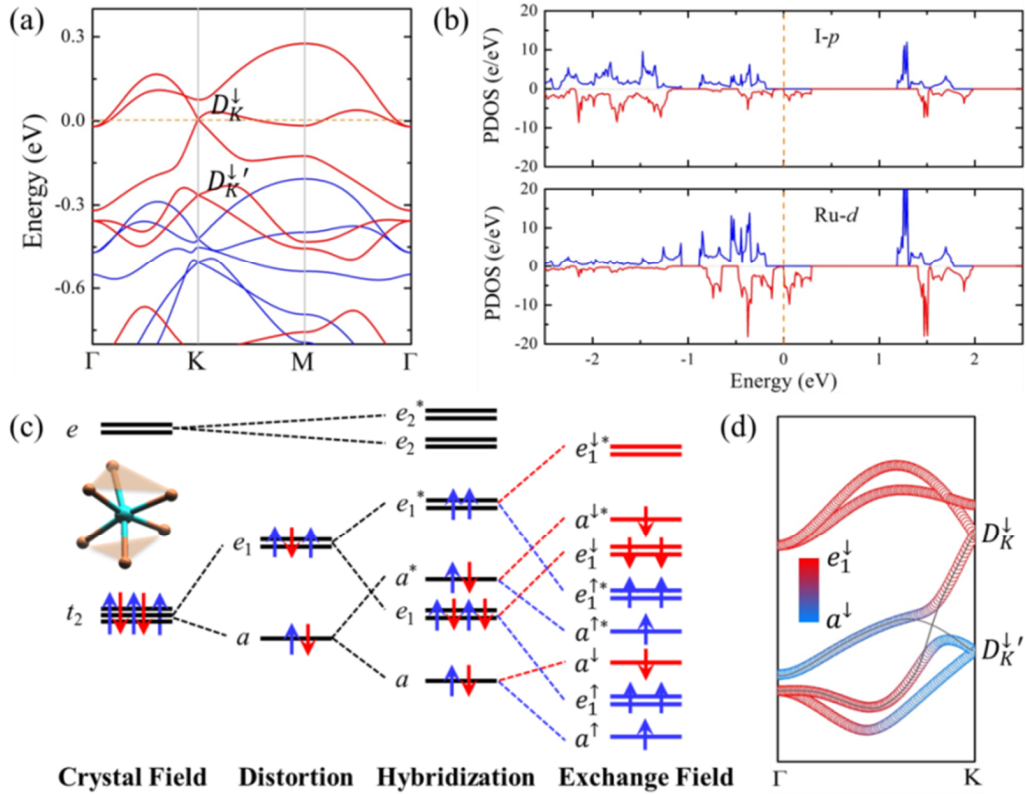

FIG. 2. (a) Band structure without SOC. Blue and red curves represent spin up and spin down bands, respectively. (b) Projected density of states. (c) Schematic diagram of the evolution from the atomic $d$ orbitals to the final states at the $\Gamma$ point. (d) Orbital-resolved spin-down bands around the Fermi level. Different colors represent proportional contribution of $e_1^\downarrow$ states and $a^\downarrow$ states. Thin black curves show the evolution of $e_1^\downarrow$ and $a^\downarrow$ states from $\Gamma$ to $K$.

To better understand the band structure, we start from the $d$ orbitals of a Ru atom [Fig. 2(c)]. Geometrically speaking, each Ru atom is coordinated by six I atoms, forming a distorted octahedral crystal field. In a perfect octahedral crystal field, the five $d$ orbitals split into $e$ and $t_2$ sub-states. In this distorted octahedral crystal field, the $t_2$ further splits into $a$ and $e$. Hence, the five $d$ orbitals split into three distinct sub-states, i.e. $a$, $e_1$, and $e_2$ (little group of Γ point is $D_{3d}$). Due to strong ligand field effect, the five $Ru^{3+}$ $d$ electrons occupy only the $a$ and $e_1$ orbitals, leaving the $e_2$ empty (distortion step). Furthermore, the hybridizations between two Ru-$a$ and Ru-$e_1$ orbitals form bonding and antibonding states. In this way, $a$, $e_1$, and, $a^*$ orbitals are fully occupied by eight electrons (four spin up and four spin down), and the degenerate $e_1^*$ states are half-filled by two spin up electrons, in keeping with the Hund's rule (hybridization step). Such half-filling also implies a stable electron configuration. The exchange between two $e_1^*$ orbitals also explains the FM ground state with a magnetic moment of 2 $\mu_B$ in one unit cell. After incorporating the magnetic exchange field, an energy split occurs between the spin up and spin down orbitals. Hence, the $a^{\downarrow*}$ and $e_1^{\downarrow*}$ lie higher in energy than the $e_1^{\uparrow*}$. This is consistent with the DFT calculated band alignments at the Γ point. Considering the honeycomb lattice of Ru atoms (which contains inversion symmetry of Ru sub-lattice), the $e_1^{\downarrow}$ and $e_1^{\downarrow*}$ bands disperse in the momentum space and form Dirac point $D_K^{\downarrow}$ at the $K$ point. Similarly, the dispersion of $a^{\downarrow}$ and $a^{\downarrow*}$ forms the $D_K^{\downarrow\prime}$ point. Thus, these Dirac points are protected by crystal symmetry of the Ru sub-lattice, and are robust against perturbations (such as in-plane strains) which keep its symmetry.

**C. SOC induced quantum anomalous Hall effect**

Now we turn on the SOC interaction. Since the system has inversion symmetry, there will be no Rashba SOC effect. Because the two bands forming $D_K^{\downarrow}$ are contributed by the same irreducible group representation ($e$), one expects that including intra-atomic SOC (**L**·**S**) would open a large local band gap. Figure 3(a) shows the band structure including SOC, where the degeneracy of $D_K^\downarrow$ is lifted, opening a direct band gap of 103 meV at $K$ and a global indirect band gap of 11 meV at the Fermi level. Similar band opening also occurs in the $D_{K'}^\downarrow$ point. Such band gap opening suggests a topologically nontrivial feature at the Fermi level. The out-of-plane spin component $\langle s_z \rangle$ of valence band is slightly quenched. In order to identify its topological property, we calculate the Berry curvature ($\Omega$) and Chern number ($C$) of each band using Kubo formula [46,47],

$$C = \sum_{n\in\{O\}} C_n = \frac{1}{2\pi} \int \sum_{n\in\{O\}} \Omega_n(k)\, d^2k = \sum_{n\in\{O\}} (C_{n,\uparrow} + C_{n,\downarrow}),$$

$$\Omega(k) = \sum_{n\in\{O\}} \Omega_n(k) = -2 \sum_{n\in\{O\}} \sum_{n'\neq n} \frac{\mathrm{Im}\,\langle\psi_{n,k}|v_x|\psi_{n',k}\rangle\langle\psi_{n',k}|v_y|\psi_{n,k}\rangle}{\left(E_{n,k} - E_{n',k}\right)^2},$$

where $n$ is the band index, $\psi_{n,k}$ is the eigenstate, $v_{x,y}$ is the velocity operator, and $\{O\}$ refers to occupied band set. The calculated Chern number of each frontier band is indicated in Fig. 3(a). The $k$-resolved Berry curvature is shown in Fig. 3(b). One clearly sees pronounced positive peaks located at $K$. Hence, the integration of Berry curvatures for all occupied bands yields a nonzero Chern number $C = -1$, indicating a quantized Hall conductance $\sigma_{xy} = C \cdot e^2/h$ within the bulk band gap. Thus, we demonstrate that the RuI$_3$ monolayer is a QAH insulator. To be specific, we adjust the chemical potential (relative to the Fermi level) and calculate the anomalous Hall conductance variation, as shown in the right panel of Fig. 3(a). We find a quantized platform of $\sigma_{xy}$ ($-1 \times e^2/h$) within the energy window of the global band gap (11 meV); $\sigma_{xy}$ gradually decreases when the chemical potential is shifted out of the band gap. Note that the $\sigma_{xy}$ remains nonzero when the chemical potential lies between $-0.1$ and $+0.2$ eV relative to $E_F$. This large range of nonzero $\sigma_{xy}$ is different from previous studies where $\sigma_{xy}$ decreases to zero rapidly out of the energy gap [12,17,48-50]. This would enhance the possibility to observe anomalous Hall conductance in experiments.

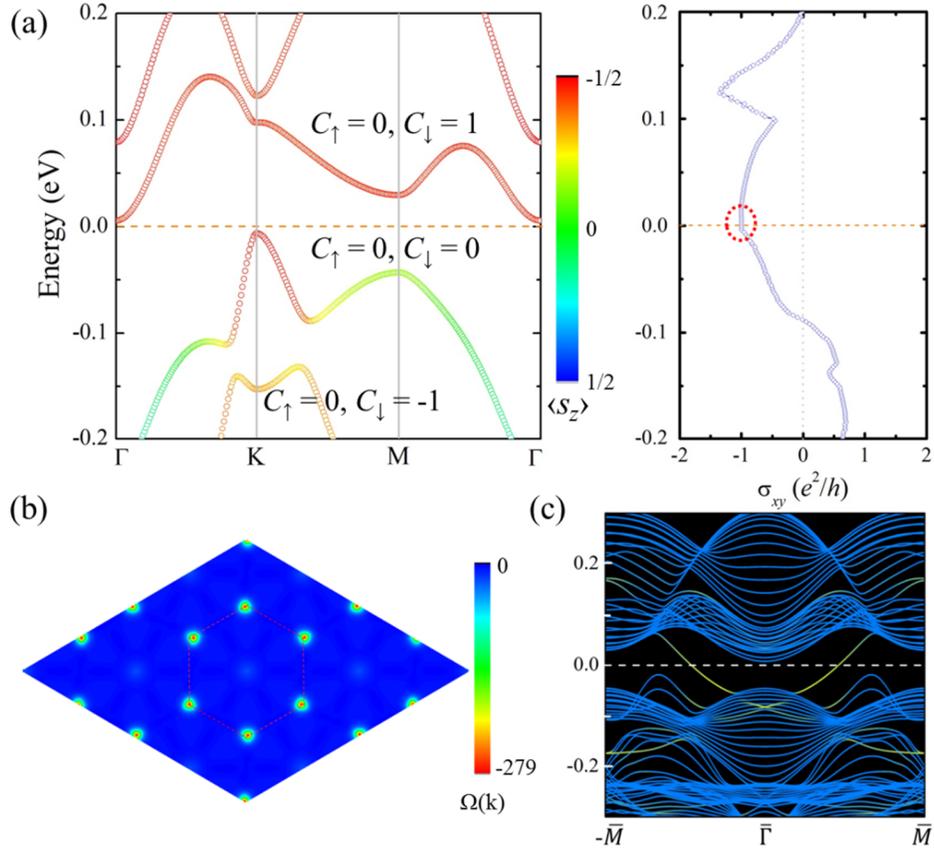

FIG. 3. (a) Band structure with SOC (left panel) and anomalous Hall conductance as a function of relative chemical potential (right panel). Different colors in the band structure represent the $\langle s_z \rangle$. The Chern numbers of frontier bands are indicated. The quantized terrace of $\sigma_{xy}$ is highlighted by the red dashed oval. (b) $k$-resolved Berry curvature $\Omega(k)$. Red dashed hexagon denotes the first Brillouin zone. (c) TB band structure of nanoribbon obtained by MLWFs show edge states (yellow) inside the gap of bulk bands (blue).

One can also confirm the QAH effect by calculating its chiral edge state within the nontrivial band gap. We fit a tight-binding Hamiltonian by using maximally localized Wannier functions to the DFT calculated bands, as implemented in the Wannier90 package. As shown in Figure S2 [39], they show very good agreement around the Fermi energy. Without loss of generality, we build a zigzag edged nanoribbon and calculate its band structure using the tight-binding Hamiltonian [Fig. 3(c)]. One clearly sees a metallic edge state appearing in the $\bar{\Gamma} \to \bar{M}$ path (the metallic state in the $-\bar{M} \to \bar{\Gamma}$ path corresponds to the opposite edge of the nanoribbon). Since the Chern number $C$ equals to the number of metallic edge states cutting the Fermi level, here we verify that $|C| = 1$.

Motivated by the recent experimental advances in magneto-optical measurement [51] we calculate its optical Hall conductivity to study how the QAH effect evolves in the ac regime. This has not been very well studied in previous computational works and should facilitate the experimental work in the future. The optical Hall conductivity can be written as

$$\sigma_{ac}(\omega) = \frac{e^2}{h} \int \frac{d^2k}{2\pi} \sum_{n' \neq n} (f_{n,k} - f_{n',k}) \frac{\text{Im}\langle \psi_{n,k}|v_x|\psi_{n',k}\rangle \langle \psi_{n',k}|v_y|\psi_{n,k}\rangle}{(\omega_{n',k} - \omega_{n,k}) - (\omega + i\eta)},$$

where $f_{n,k}$ is Fermi-Dirac distribution, $\omega$ is incident optical frequency, and $\eta$ is an infinitesimal parameter. By tuning the chemical potential, we plot the real and imaginary parts of $\sigma_{ac}$ (Fig. 4), which reflect the reactive and dissipative behavior of an incident photon, respectively. We observe that $\sigma_{ac}$ strongly fluctuates when $0 < \hbar\omega < 0.5$ eV and $1.3 < \hbar\omega < 2.5$ eV. It almost diminishes when $\hbar\omega$ lies in the range 0.5 to 1.3 eV, which is mainly due to the large gap between the 0.3 and 1.2 eV in the band structure (Fig. S2 in [39]). In the dc limit ($\omega = 0$), the real part of $\sigma_{ac}$ is essentially identical to $\sigma_{xy}$. The real and imaginary parts of $\sigma_{ac}$ in the intrinsic state (chemical potential at the Fermi level) are also shown in Fig. S3 [39]. Although the $\sigma_{ac}$ shows very complex structure, there are still some features to be noticed. Naively, one expects that the terrace in the optical conductivity immediately vanishes in the ac regime as there is no topological protection. However, in the intrinsic state, the real part of $\sigma_{ac}$ is around $-1$ $e^2/h$ up to $\hbar\omega = 0.1$ eV. This would help the experimental observation of large anomalous Hall effect. In the $p$-doping state (negative relative chemical potential), one always sees a large ac Hall plateau of ~2 $e^2/h$ at $\hbar\omega = 0.4$ eV, which disappears in the $n$-doping state.

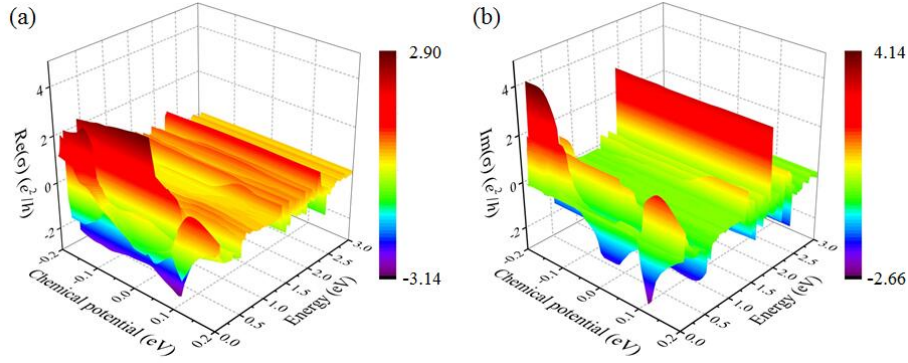

FIG. 4. (a) Real and (b) imaginary parts of the optical conductivity σ$_{xy}$ with respect to photonic energy and chemical potential.

**D. In-plane strain effect**

In order to further study the QAH effects of RuI$_3$, we calculate the in-plane strain effect on magnetic exchange and the global band gap (Fig. 5). We find that, with the nontrivial band topology preserved, a compressive strain increases the bulk band gap, while the tensile strain decreases it. The nontrivial band gap becomes 21 meV when a 2% in-plane compression is applied. On the other hand, the exchange parameter $J$ increases monotonically as the lattice expands. Thus, one can apply an appropriate in-plane strain to achieve an optimal working temperature.

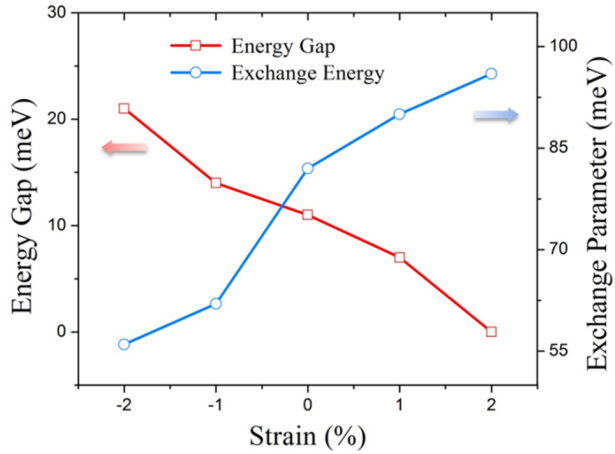

FIG. 5. Nontrivial bulk energy gap and magnetic exchange parameter $J$ as functions of biaxial in-plane strain.

### E. Ferromagnetic RuCl₃ and RuBr₃ monolayers

Besides RuI$_3$ monolayer we also investigate similar RuCl$_3$ and RuBr$_3$ monolayers. Our GGA and GGA+$U$ calculations show that their exchange energies are very small and sensitive to the effective $U$ values. This suggests that the ground states of RuCl$_3$ and RuBr$_3$ monolayers lie at the border between FM and AFM configurations. Hence, in order to achieve robust FM states, one needs to explicitly apply a weak external magnetic field or a small in-plane strain. Nevertheless, we also find similar topological features in FM RuCl$_3$ and RuBr$_3$ monolayers. The calculated band structures of FM RuCl$_3$ and RuBr$_3$ monolayers show similar behavior as the RuI$_3$ monolayer (Fig. 6). When the SOC is absent, we again find Dirac point at the $K$ point in the spin down channel. The SOC lifts the degeneracy of Dirac point and a nontrivial energy gap opens at $K$, showing the same nontrivial band topology as in RuI$_3$ monolayer. However, in both cases, the conduction band drops below the Fermi level around the $\Gamma$ point, and the valence band lies above the Fermi level around the $M$ point. Thus both of these materials would show semi-metallic features rather than QAH insulating. In spite of this, due to our previous results for RuI$_3$ monolayer, one still would observe chiral dissipation-less edge state in their corresponding nanoribbons, and expect that the QAH insulating state can be achieved by applying a weak external magnetic field and/or small in-plane strain.

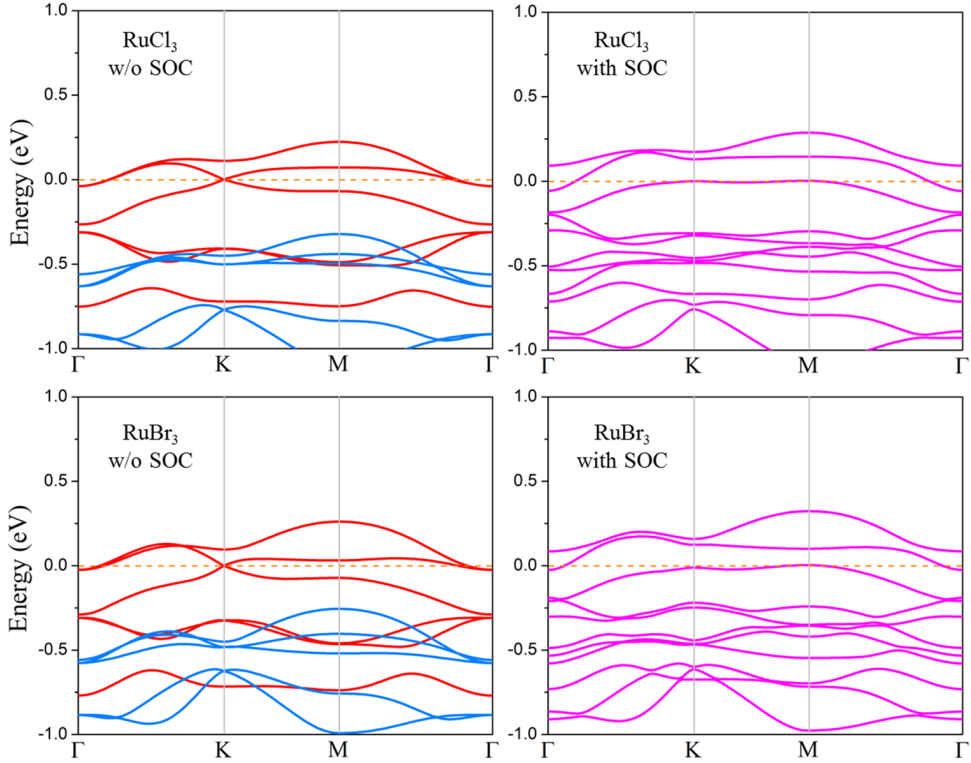

FIG. 6. Band structures of ferromagnetic states for RuCl$_3$ and RuBr$_3$ monolayer without and with SOC. Blue and red curves in left panels denote spin up and spin down channels, respectively.

**IV. Conclusion**

In summary, based on first-principles DFT calculations, we predict that 2D RuI$_3$ monolayer is an intrinsic FM QAH insulator. This material could be synthesized in experiments and the TM atoms are uniformly distributed. The Curie temperature is estimated to be ~360 K, higher than most of the 2D FM thin films studied hitherto. Without including SOC interaction, a Dirac point in the spin down channel appears at the Fermi level, which is contributed by Ru-$d$ orbitals and protected by crystal symmetry of Ru sublattice. The mechanism of such Dirac point has been understood by considering the ligand field effect, hybridization, and magnetic exchange field interactions. Once the SOC is introduced, the symmetry protected Dirac point opens a band gap and the system becomes a QAH insulator with a global band gap of 11 meV. Thus, the topologically nontrivial band gap is robust against perturbations that retain its crystal symmetry. The nontrivial band topology and intrinsic QAH effect are demonstrated by calculating its Berry curvature, Chern number, and chiral edge state. In-plane strain effects are also discussed which are expected to play a role in tailoring both the band gap and the Curie temperature. We look forward to experimental verifications of the QAH effects in the ruthenium halide family.

**Acknowledgement.**

E.K. is supported by the NSFC (11374160, 51522206, 11574151), by NSF of Jiangsu Province (BK20130031), by PAPD, the Fundamental Research Funds for the Central Universities (No.30915011203), and by New Century Excellent Talents in University (NCET-12-0628). C.H. and E.K. acknowledge the support from the Shanghai Supercomputer Centre. P.J. acknowledges support by the U.S. Department of Energy, Office of Basic Energy Sciences, Division of Materials Sciences and Engineering under Award # DE-FG02-96ER45579. J.Z. and P.J. acknowledge the resources of the National Energy Research Scientific Computing Center supported by the Office of Science of the U.S. Department of Energy under Contract no. DE-AC02-05CH11231. C.H. acknowledges the China Scholarship Council (CSC) for sponsoring his visit to Virginia Commonwealth University (VCU) where this work was conducted.

C.H. and J.Z. contributed equally to this paper.

**References:**

[1] J. E. Moore, Nature **464**, 194 (2010).

[2] M. Z. Hasan and C. L. Kane, Rev. Mod. Phys. **82**, 3045 (2010).

[3] X.-L. Qi and S.-C. Zhang, Phys. Today **63**, 33 (2010).

[4] Y. Ren, Z. Qiao, and Q. Niu, Rep. Prog. Phys. **79**, 066501 (2016).

[5] J. G. Checkelsky, R. Yoshimi, A. Tsukazaki, K. S. Takahashi, Y. Kozuka, J. Falson,

M. Kawasaki, and Y. Tokura, Nat. Phys. **10**, 731 (2014).

[6] H. Weng, R. Yu, X. Hu, X. Dai, and Z. Fang, Adv. Phys. **64**, 227 (2015).

[7] C.-X. Liu, S.-C. Zhang, and X.-L. Qi, Annu. Rev. Condens. Matter Phys. **7**, 301 (2016).

[8] F. D. M. Haldane, Phys. Rev. Lett. **61**, 2015 (1988).

[9] R. Yu, W. Zhang, H. J. Zhang, S. C. Zhang, X. Dai, and Z. Fang, Science **329**, 61 (2010).

[10] G. Xu, J. Wang, C. Felser, X.-L. Qi, and S.-C. Zhang, Nano Lett. **15**, 2019 (2015).

[11] S. Qi, Z. Qiao, X. Deng, E. D. Cubuk, H. Chen, W. Zhu, E. Kaxiras, S. B. Zhang, X. Xu, and Z. Zhang, Phys. Rev. Lett. **117**, 056804 (2016).

[12] H. Zhang, C. Lazo, S. Blügel, S. Heinze, and Y. Mokrousov, Phys. Rev. Lett. **108**, 056802 (2012).

[13] Z. Qiao, S. A. Yang, W. Feng, W.-K. Tse, J. Ding, Y. Yao, J. Wang, and Q. Niu, Phys. Rev. B **82**, 161414 (2010).

[14] H. Pan, Z. Li, C.-C. Liu, G. Zhu, Z. Qiao, and Y. Yao, Phys. Rev. Lett. **112**, 106802 (2014).

[15] M. Ezawa, Phys. Rev. Lett. **109**, 055502 (2012).

[16] Z. F. Wang, Z. Liu, and F. Liu, Phys. Rev. Lett. **110**, 196801 (2013).

[17] L. Dong, Y. Kim, D. Er, A. M. Rappe, and V. B. Shenoy, Phys. Rev. Lett. **116**, 096601 (2016).

[18] K. F. Garrity and D. Vanderbilt, Phys. Rev. Lett. **110**, 116802 (2013).

[19] C. Wu, Phys. Rev. Lett. **101**, 186807 (2008).

[20] J. Zhou, Q.-F. Liang, H. Weng, Y. B. Chen, S.-H. Yao, Y.-F. Chen, J. Dong, and G.-Y. Guo, Phys. Rev. Lett. **116**, 256601 (2016).

[21] S.-C. Wu, G. Shan, and B. Yan, Phys. Rev. Lett. **113**, 256401 (2014).

[22] C.-Z. Chang, J. Zhang, X. Feng, J. Shen, Z. Zhang, M. Guo, K. Li, Y. Ou, P. Wei, L.-L. Wang, Z.-Q. Ji, Y. Feng, S. Ji, X. Chen, J. Jia, X. Dai, Z. Fang, S.-C. Zhang, K. He, Y. Wang, L. Lu, X.-C. Ma, and Q.-K. Xue, Science **340**, 167 (2013).

[23] X. Kou, S.-T. Guo, Y. Fan, L. Pan, M. Lang, Y. Jiang, Q. Shao, T. Nie, K. Murata, J. Tang, Y. Wang, L. He, T.-K. Lee, W.-L. Lee, and K. L. Wang, Phys. Rev. Lett. **113**, 137201 (2014).

[24] C.-Z. Chang, W. Zhao, D. Kim, Y. H. Zhang, B. A. Assaf, D. Heiman, S.-C. Zhang, C. Liu, M. H. W. Chan, and J. S. Moodera, Nat. Mater. **14**, 473 (2015).

[25] D. Zhang, A. Richardella, D. W. Rench, S.-Y. Xu, A. Kandala, T. C. Flanagan, H. Beidenkopf, A. L. Yeats, B. B. Buckley, P. V. Klimov, D. D. Awschalom, A. Yazdani, P. Schiffer, M. Zahid Hasan, and N. Samarth, Phys. Rev. B **86**, 205127 (2012).

[26] J. Zhang, C.-Z. Chang, P. Tang, Z. Zhang, X. Feng, K. Li, L.-l. Wang, X. Chen, C. Liu, W. Duan, K. He, Q.-K. Xue, X. Ma, and Y. Wang, Science **339**, 1582 (2013).

[27] S. I. Troyanov and E. M. Snigireva, Zh. Neorg. Khim. **36**, 1117 (1991).

[28] H. Bengel, H. J. Cantow, S. N. Magonov, H. Hillebrechtb, G. Thieleb, W. Liang, and M. H. Whangbo, Surf. Sci. **343**, 95 (1995).

[29] A. Banerjee, C. A. Bridges, J.-Q. Yan, A. A. Aczel, L. Li, M. B. Stone, G. E. Granroth, M. D. Lumsden, Y. Yiu, J. Knolle, S. Bhattacharjee, D. L. Kovrizhin, R. Moessner, D. A. Tennant, D. G. Mandrus, and S. E. Nagler, Nat. Mater. **15**, 733 (2016).

[30] V. Nicolosi, M. Chhowalla, M. G. Kanatzidis, M. S. Strano, and J. N. Coleman, Science **340**, 1226419 (2013).

[31] J. Liu, Q. Sun, Y. Kawazoe, and P. Jena, Phys. Chem. Chem. Phys. **18**, 8777 (2016).

[32] W.-B. Zhang, Q. Qu, P. Zhu, and C.-H. Lam, J. Mater. Chem. C **3**, 12457 (2015).

[33] Y. Zhou, H. Lu, X. Zu, and F. Gao, Sci. Rep. **6**, 19407 (2016).

[34] P. Perdew, K. Burke, and M. Ernzerhof, Phys. Rev. Lett. **77**, 3865 (1996).

[35] G. Kresse and J. Hafner, Phys. Rev. B **47**, 558 (1993).

[36] P. E. Blöchl, Phys. Rev. B **50**, 17953 (1994).

[37] H. J. Monkhorst and J. D. Pack, Phys. Rev. B **13**, 5188 (1976).

[38] S. Dudarev, G. Botton, Y. Savrasov, C. Humphreys, and A. Sutton, Phys. Rev. B **57**, 1505 (1998).

[39] Supplemental Material for AIMD simulation results, different magnetic coupling configurations, comparison of MLWF and DFT calculated band structure.

[40] N. Marzari, A. A. Mostofi, J. R. Yates, I. Souza, and D. Vanderbilt, Rev. Mod. Phys. **84**, 1419 (2012).

[41] A. A. Mostofi, J. R. Yates, Y.-S. Lee, I. Souza, D. Vanderbilt, and N. Marzari, Comput. Phys. Commun. **178**, 685 (2008).

[42] H.-S. Kim, V. V. Shankar, A. Catuneanu, and H.-Y. Kee, Phys. Rev. B **91**, 241110 (2015).

[43] M. Kan, B. Wang, Y. H. Lee, and Q. Sun, Nano. Res. **8**, 1348 (2015).

[44] X. Li, X. Wu, and J. Yang, J. Am. Chem. Soc. **136**, 11065 (2014).

[45] J. Zhou and Q. Sun, J. Am. Chem. Soc. **133**, 15113 (2011).

[46] Y. G. Yao and Z. Fang, Phys. Rev. Lett. **95**, 156601 (2005).

[47] G. Y. Guo, Y. G. Yao, and Q. Niu, Phys. Rev. Lett. **94**, 226601 (2005).

[48] Q.-Z. Wang, X. Liu, H.-J. Zhang, N. Samarth, S.-C. Zhang, and C.-X. Liu, Phys. Rev. Lett. **113**, 147201 (2014).

[49] Z. Qiao, W. Ren, H. Chen, L. Bellaiche, Z. Zhang, A. H. MacDonald, and Q. Niu, Phys. Rev. Lett. **112**, 116404 (2014).

[50] P. Zhou, C. Q. Sun, and L. Z. Sun, Nano Lett. **16**, 6325 (2016).

[51] H. Sumikura, T. Nagashima, H. Kitahara, and M. Hangyo, Jpn. J. Appl. Phys. **46**, 1739 (2007).